\documentclass{elsart1p}

\usepackage{graphicx}
\usepackage{amsmath}
\usepackage{amssymb}

\begin{document}

%%%%%%%%%%%%%%%%%%%%%%%%%%%%%%%%%%%%%%%%%%%%%%%%%%%%%%%%%%%%%%%%%%%%%%%%%%%%%%%

\begin{frontmatter}

% Title, authors and addresses

% use the thanksref command within \title, \author or \address for footnotes;
% use the corauthref command within \author for corresponding author footnotes;
% use the ead command for the email address,
% and the form \ead[url] for the home page:
% \title{Title\thanksref{label1}}
% \thanks[label1]{}
% \author{Name\corauthref{cor1}\thanksref{label2}}
% \ead{email address}
% \ead[url]{home page}
% \thanks[label2]{}
% \corauth[cor1]{}
% \address{Address\thanksref{label3}}
% \thanks[label3]{}

% use optional labels to link authors explicitly to addresses:
% \author[label1,label2]{}
% \address[label1]{}
% \address[label2]{}

%%%%%%%%%%%%%%%%%%%%%%%%%%%%%%%%%%%%%%%%%%%%%%%%%%%%%%%%%%%%%%%%%%%%%%%%%%%%%%%

\title{The Tensor Part of the Skyrme Energy Density Functional}

%%%%%%%%%%%%%%%%%%%%%%%%%%%%%%%%%%%%%%%%%%%%%%%%%%%%%%%%%%%%%%%%%%%%%%%%%%%%%%%

\author[IPNL]{T. Lesinski\corauthref{TL}}
\author[SPhN,CENBG]{M. Bender\thanksref{ESNTthks}\corauthref{MB}}
\author[IPNL,SPhN]{K. Bennaceur\thanksref{ESNTthks}}
\author[NSCL]{T. Duguet}
\author[IPNL]{J. Meyer}

%%%%%%%%%%%%%%%%%%%%%%%%%%%%%%%%%%%%%%%%%%%%%%%%%%%%%%%%%%%%%%%%%%%%%%%%%%%%%%%

\address[IPNL]{Universit{\'e} de Lyon, F-69003 Lyon, France;
             Institut de Physique Nucl{\'e}aire de Lyon,
             CNRS/IN2P3, Universit{\'e} Lyon 1,
             F-69622 Villeurbanne, France}
\address[SPhN]{DSM/DAPNIA/SPhN, CEA Saclay,
             F-91191 Gif-sur-Yvette Cedex,
             France}
\address[CENBG]{Universit{\'e} Bordeaux 1; CNRS/IN2P3;
             Centre d'{\'E}tudes Nucl{\'e}aires de Bordeaux Gradignan, UMR5797,
             Chemin du Solarium, BP120, F-33175 Gradignan, France}
\address[NSCL]{National Superconducting Cyclotron Laboratory
             and Department of Physics and Astronomy,
             Michigan State University, East Lansing, MI 48824,
             USA}

\corauth[TL]{email: lesinski@ipnl.in2p3.fr}
\corauth[MB]{email: bender@cenbg.in2p3.fr}

\thanks[ESNTthks]{Work performed in the framework of the \textit{Espace de
Structure Nucleaire Th{\'e}orique}.}

%%%%%%%%%%%%%%%%%%%%%%%%%%%%%%%%%%%%%%%%%%%%%%%%%%%%%%%%%%%%%%%%%%%%%%%%%%%%%%%

\begin{abstract}
We systematically study the effect of the $\vec{J}^2$ tensor terms in
the Skyrme energy functional on properties of spherical nuclei. We build a set
of 36 parameterizations covering a wide range of the corresponding parameter
space. We analyze the impact of the tensor terms on the evolution of
single-particle-level splittings along chains of semi-magic nuclei in spherical
calculations. We find that positive values of the coupling constants of
proton-neutron and like-particle tensor terms allow for a qualitative
description of the evolution of neutron and proton single-particle level
splittings in chains of Ca, Ni and Sn isotopes.
%For the same values of the
%tensor term coupling constants, however, the overall agreement of the
%single-particle spectra in doubly-magic nuclei is deteriorated, which can be
%traced back to the currently used central and spin-orbit parts of the Skyrme
%energy density functional. These appear not flexible enough to allow for the
%presence of large tensor terms.
\end{abstract}

%%%%%%%%%%%%%%%%%%%%%%%%%%%%%%%%%%%%%%%%%%%%%%%%%%%%%%%%%%%%%%%%%%%%%%%%%%%%%%%

\end{frontmatter}

%%%%%%%%%%%%%%%%%%%%%%%%%%%%%%%%%%%%%%%%%%%%%%%%%%%%%%%%%%%%%%%%%%%%%%%%%%%%%%%
%%%%%%%%%%%%%%%%%%%%%%%%%%%%%%%%%%%%%%%%%%%%%%%%%%%%%%%%%%%%%%%%%%%%%%%%%%%%%%%

The tensor force has been identified very early as an important part of the
nucleon-nucleon interaction, and its effects, e.g. the specific correlations it
generates and their importance for the binding of nuclear systems, have been
studied in infinite and few-body systems. However, it is only recently that
energy density functional (EDF) practitioners have renewed their interest in the
various ``tensor terms'' occuring in the nuclear EDF
\cite{Ots05a,Ots06a,Dobaczewskitalk,Lon06b,Bro06a,Col07a,Bri07a}.
%Whereas these
%references mainly aim at pinning down a ``best'' parameterization of these
%tensor terms,
We attempt in this contribution to sketch a systematic study of the variation of
tensor-term parameters and constrain their values \cite{Les07a}.

In spherical symmetry, a zero-range tensor force added on top of the usual
Skyrme effective vertex \cite{RMP} contributes to the EDF through terms
proportional to $\vec{J}^2$ \cite{Sta77a,Per04a} where $\vec{J}$ is the
spin-orbit current density vector, here defined through its radial component:
\begin{eqnarray}
\label{eq:j:radial}
J_q (r)
& = & \frac{1}{4 \pi r^3} \! \sum_{n,j,\ell}
       ( 2 j + 1 ) \, v_{n j \ell}^2
      \;
      \Big[  j ( j + 1)
           - \ell ( \ell + 1)
           - \tfrac{3}{4} \Big] \, \psi^2_{n j \ell} (r).
\end{eqnarray}
The resulting total spin-orbit field for neutrons reads (invert n and p for
protons)
\begin{eqnarray}
\label{eq:wpot:tot}
W_n (r)
& = & \frac{W_0}{2} \, \big( 2\nabla \rho_n + \nabla \rho_p )
      + \alpha\, J_n + \beta \,J_p \,,
\end{eqnarray}
where the first term comes from the zero-range spin-orbit vertex and the two
others from the tensor vertex. When the functional is derived from such a
Skyrme-tensor vertex the coupling parameters $\alpha$ and $\beta$ can be chosen
independently of the more standard force or functional parameters. In
order to study their effects, we build a series of parameterizations, for each
of which $(\alpha,\beta)$ are fixed and all other parameters are fitted
according to a protocol~\cite{Les07a} similar to the one used for the
construction of the Saclay-Lyon parameterizations. They are labelled T$IJ$,
with indices $I$ and $J$ related to $\alpha$ and $\beta$ through $\alpha =
60\,(J-2)\,$MeV\,fm$^5$ and $\beta = 60\,(I-2)\,$MeV\,fm$^5$.

% \begin{figure}[b]
%   \centering
%   \includegraphics[width=0.5\columnwidth]{cjplane-tot.eps}
%   \caption{
%     Values of $C^{J}_0$ and $C^{J}_1$ for our set of parameterizations
%     (circles). Diagonal lines indicate $\alpha = C^{J}_0 + C^{J}_1 = 0$
%     (pure neutron-proton coupling) and $\beta = C^{J}_0 - C^{J}_1 = 0$ (pure
%     like-particle coupling). Values for classical parameter sets are also
%     indicated (dots), with SLy4 representing all parameterizations
%     for which $\vec J^2$ terms have been omitted in the fit. Recent
%     parameterizations with tensor terms are indicated by squares.
%   }
%   \label{fig:cjplane-tot}
% \end{figure}

Tensor terms alter the strength and shape of the spin-orbit potential,
Eq.~(\ref{eq:wpot:tot}), when $\vec{J}$ varies due to the filling of a
single-particle state. The tensor contribution to $W_q(r)$ thus depends
on details of the relative placement of the levels, and is subject to much
sharper relative variations than the spin-orbit contribution. As such, it can
be constrained by examining the variation of the relative placement of
single-particle states in a series of nuclei differing by the filling of levels
which significantly contribute to $\vec{J}$.

\begin{figure}[t]
  \centering
  \includegraphics[width=0.5\columnwidth,clip]
{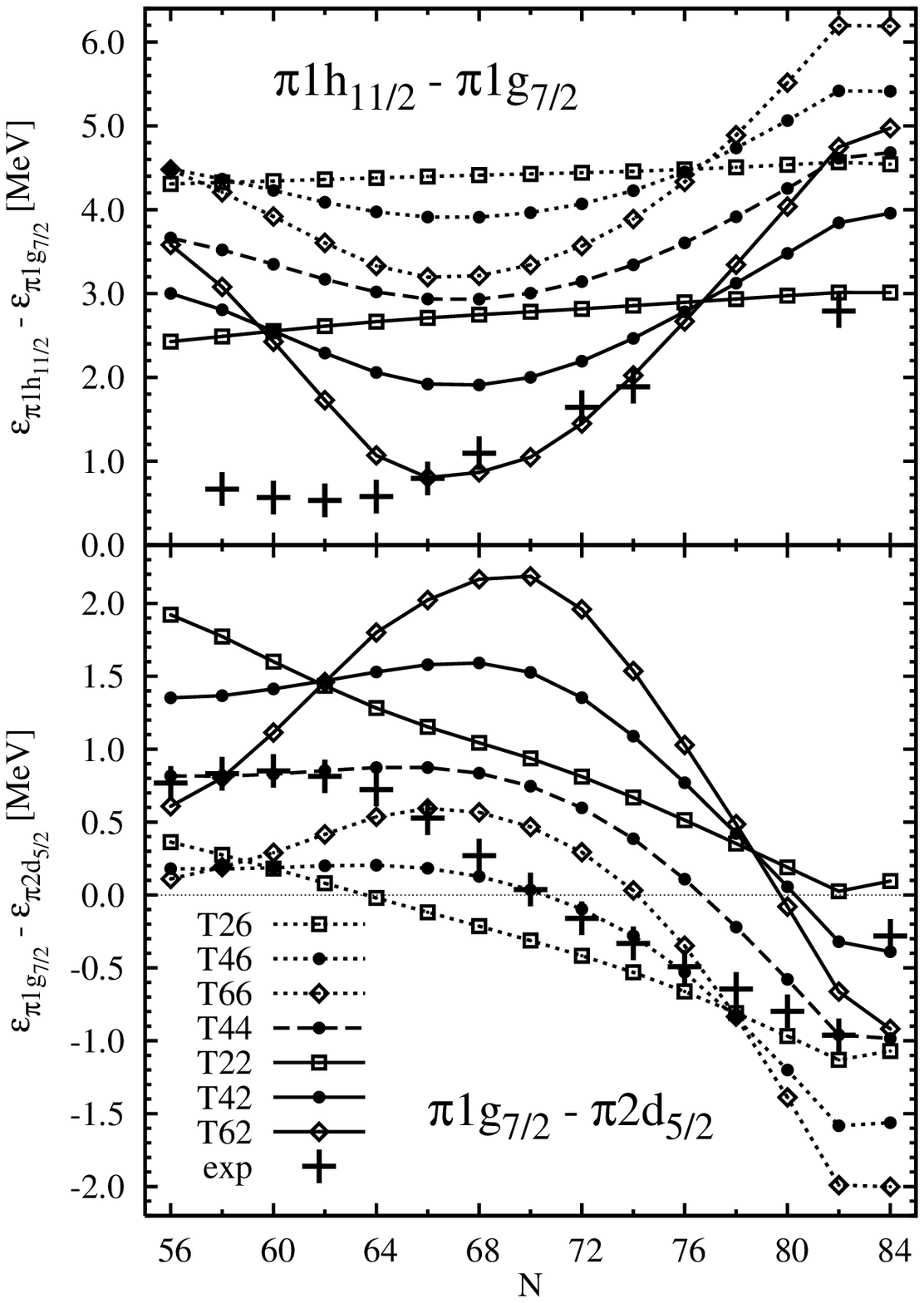}
  \includegraphics[width=0.4\columnwidth]{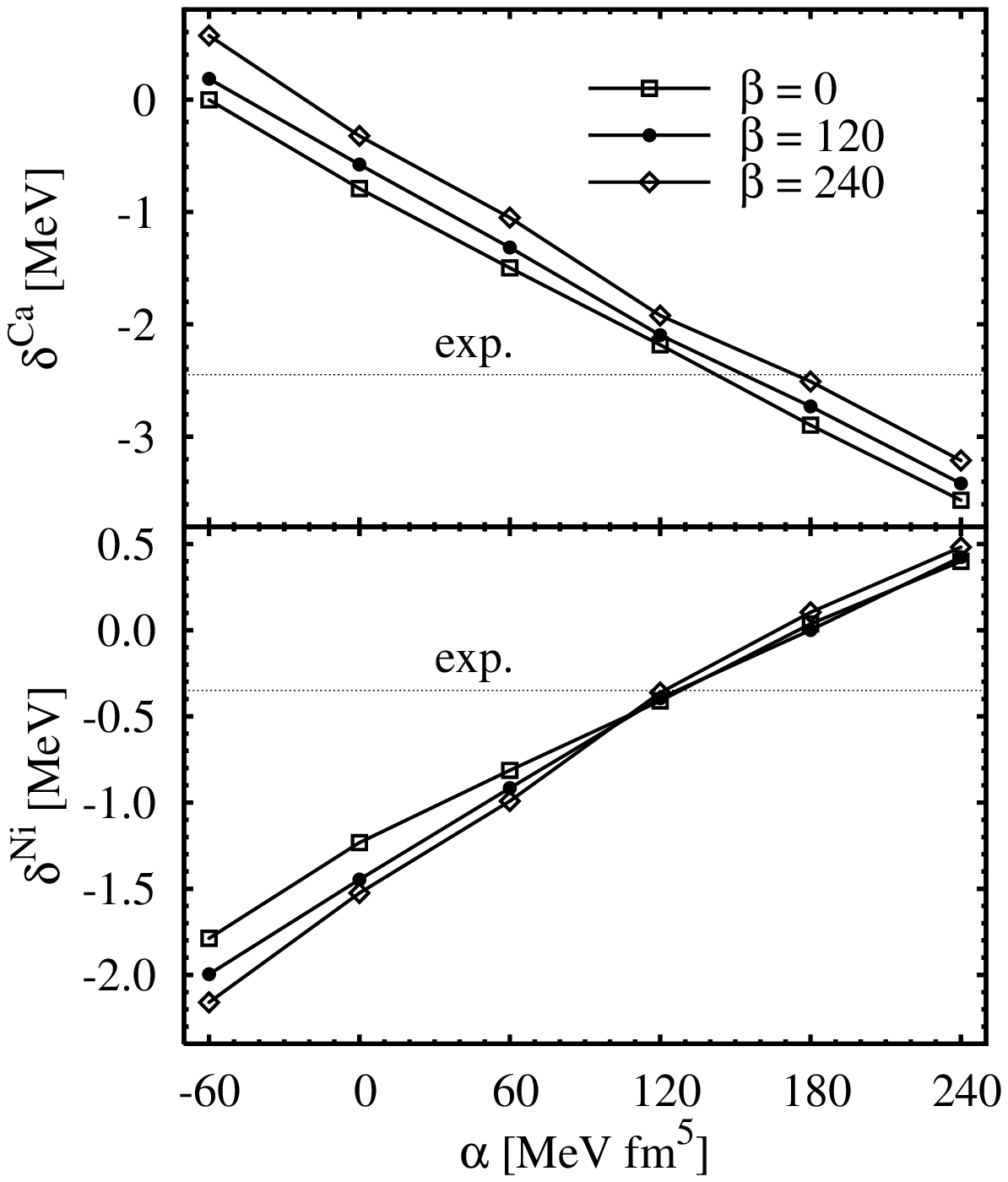}
  \caption{
    Incidence of a variation of $\vec{J}^2$ coupling constants on
    single-particle level shifts.
    Left panel: Distance of the proton $1h_{11/2}$ and $1g_{7/2}$ levels (top)
    and of the proton $2d_{5/2}$ and $1g_{7/2}$ levels (bottom), for the
    chain of tin isotopes.
    % The ``best'' parameterization cannot and should
    % not be determined with a $\chi^2$ criterion, see text.
    Right panel: Shift of the distance between the neutron $1d_{3/2}$
    and $2s_{1/2}$ levels when going from  \nuc{40}{Ca} to \nuc{48}{Ca},
    Eq.~(\ref{eq:nnshift-Ca}) (top) and of the neutron $1f_{5/2}$ and
    $2p_{1/2}$ levels when going from \nuc{56}{Ni} to \nuc{68}{Ni},
    Eq.~(\ref{eq:nnshift-Ni}) (bottom).
  }
  \label{fig:spe-shift}
\end{figure}

The first example, displayed on the left panel of Fig.~\ref{fig:spe-shift}, is
the tin chain, along which the $h_{11/2}$ neutron level is filled (between
$N=64$ and $82$) yielding a large contribution to $\vec{J}_\text{n}$ and thus
to the proton spin-orbit field due to the
$\beta\,\vec{J}_\text{n}\cdot\vec{J}_\text{p}$ coupling. This has been
previously identified as a possible source of the change of slope (as a function
of $N$) in the spacing of proton $1g_{7/2}$ and $1h_{11/2}$ levels
\cite{Schi04aE}. The spacing of $2d_{5/2}$ and $1g_{7/2}$ levels is affected in
a similar way. We can reproduce the magnitude of the single-particle-level
spacing shifts by setting the np-coupling $\beta$ to $120\,$MeV\,fm$^5$,
however, the effect appears to occur at too large a neutron number, owing to
the incorrect placement of the neutron $1h_{11/2}$ level relative to the
$3s_{1/2}$ and $2d_{3/2}$ ones.

The neutron-neutron coupling can be constrained by examining the spacings of
neutron levels in nuclei of the same isotopic chain. In the Ca chain, for
example, the filling of the $1f_{7/2}$ level between $^{40}$Ca and $^{48}$Ca
affects the splitting of the neutron $1d$ shell, yielding a shift of the
$1d_{3/2}$ relative to the $2s_{1/2}$ one. Similarly, the filling of the
$1f_{5/2}$ level between $^{56}$Ni and $^{68}$Ni acts on the $2p$ and $1f$
states and produces a relative shift of the $1f_{5/2}$ and $2p_{1/2}$ levels.
The right panel of Fig.~\ref{fig:spe-shift} displays the evolution of level
splittings related to the latter effects, as a function of the like-particle
coupling constant $\alpha$:
\begin{eqnarray}
\delta^\text{Ca}
&=& \left(   \varepsilon_{1d_{3/2}}^{\text{\nuc{48}{Ca}}}
         - \varepsilon_{2s_{1/2}}^{\text{\nuc{48}{Ca}}}
    \right)
  - \left(   \varepsilon_{1d_{3/2}}^{\text{\nuc{40}{Ca}}}
           - \varepsilon_{2s_{1/2}}^{\text{\nuc{40}{Ca}}}
    \right),
\label{eq:nnshift-Ca}
    \\
\delta^\text{Ni}
&=& \left(   \varepsilon_{1f_{5/2}}^{\text{\nuc{68}{Ni}}}
         - \varepsilon_{2p_{1/2}}^{\text{\nuc{68}{Ni}}}
    \right)
  - \left(   \varepsilon_{1f_{5/2}}^{\text{\nuc{56}{Ni}}}
           - \varepsilon_{2p_{1/2}}^{\text{\nuc{56}{Ni}}}
    \right).
\label{eq:nnshift-Ni}
\end{eqnarray}
The two cases are consistent with each other, as a satisfactory comparison to
experiment is obtained for values in the range
$\alpha\simeq 120-150\,$MeV\,fm$^5$.

To conclude, we have constrained the tensor-term parameters, yielding
$\alpha\sim\beta\sim 120$\,MeV\,fm$^5$. However, the addition of these terms
does not provide a global improvement of single-particle spectra, as has been
expected from more limited studies, and even deteriorates some aspects of
single-particle spectra in doubly-magic nuclei \cite{Les07a}. Complementary work
on the central and spin-orbit parts should thus be performed.

%\nocite{Ben06a, Dob84a,Sky56a,Bel56a,Cha97a,Cha98a,
%Ben05a,  Ben99b, Ber80a, Lit06a, Dob94a, Bei75a}

%%%%%%%%%%%%%%%%%%%%%%%%%%%%%%%%%%%%%%%%%%%%%%%%%%%%%%%%%%%%%%%%%%%%%%%%%%%%%%%
%%%%%%%%%%%%%%%%%%%%%%%%%%%%%%%%%%%%%%%%%%%%%%%%%%%%%%%%%%%%%%%%%%%%%%%%%%%%%%%

%\bibliographystyle{elsart-num}
%\bibliography{tensor}

%%%%%%%%%%%%%%%%%%%%%%%%%%%%%%%%%%%%%%%%%%%%%%%%%%%%%%%%%%%%%%%%%%%%%%%%%%%%%%%
%%%%%%%%%%%%%%%%%%%%%%%%%%%%%%%%%%%%%%%%%%%%%%%%%%%%%%%%%%%%%%%%%%%%%%%%%%%%%%%

\end{document}